\documentstyle{article}
\begin{document}
\begin{center}
\large
\bf
Neuro-flow Dynamics and the Learning Processes\\
\rm
\vskip 0.3cm
\normalsize
{\bf M. TATSUNO and Y. AIZAWA}\\
\vskip 0.3cm
Department of Applied Physics, School of Science and Engineering\\
Waseda University, Shinjuku, Tokyo 169, JAPAN\\
\end{center}
{\bf Abstract} -- A new description of the 
neural activity is introduced by the neuro-flow dynamics
and the extended Hebb rule.  The remarkable 
characteristics of the neuro-flow dynamics, such as the
primacy and the recency effect during awakeness or 
sleep, are
pointed out.
\begin{center}
(Accepted 2 December 1996)
\end{center}
\begin{center}
{\bf 1. INTRODUCTION}\\
\end{center}
Research of the central nervous systems has attracted much
attention recently, and a large number of contributions,
especially theoretical ones, have been made by physicists
since the 1980's\cite{la1,la2,la3}.  
In these researches, theoreticians have made very simplified 
models of the brain, and studied their statistical properties
analytically as well as numerically.
Among those
models, the Hopfield model and its extensions
have been throughly studied, and many interesting properties
have been
obtained, including the famous
$p/N \approx 0.14$ limit where $N$ is
the number of neurons and $p$ is the number of stored patterns \cite{la2,la4}.
An important hypothesis generally used on these models is the Hebb
rule,
that is to say, ``the changes of synaptic strength are
proportional to the correlation between the firing of
the pre- and post-synaptic neurons'' \cite{la5}.
The original idea by Hebb has been changed a little,
i.e., firing of the post-synaptic neuron is not
always necessary and the change of the membrane potential 
is more significant for synaptic plasticity \cite{la6,la7}.
However there still remain many problems in the
molecular process of synaptic transmission.  
The molecular basis of the Hebb rule has been pursued
precisely to explain synaptic activity \cite{la7}.\\
\indent
Many qualitative behaviors of neural networks have been elucidated
based on the Hebb rule mentioned above, but there
still remain many differences between the real
functions of the brain
and artificial neural networks.  The most important of them
is that the real brain always works in the non-equilibrium
states accompanied by nerve impulses.  On the other hand,
models of artificial neural networks have been treated
in the equilibrium state of the thermo-dynamical limit.
Thus, it is inevitable that the model of neural networks is
extended to the
non-equilibrium state.  In this paper, we will propose a new
model of neuro-flow dynamics in which the essential variable
to describe the neural activity is the flow rate
of neuro-impulses which propagate from the $j$th to the $i$th
neuron.  Furthermore, the original Hebb rule should be
extended to include synaptic plasticity due to the neural flow.
In this paper we will discuss some remarkable effects generated by the
neuro-flow dynamics with the extended Hebb rule.\\ 
\begin{center}
{\bf 2. MODEL OF NEURO-FLOW DYNAMICS}\\ 
\end{center}
\noindent
The standard representation of neural networks 
in
discrete time is written by
\begin{equation}
\sigma_{i}(t+1)=f_{T}(\sum_{j=1}^{N}w_{ij}(t)\sigma_{j}(t)-h_{i})
\end{equation}
\noindent
where $\sigma_{i}$ and $h_{i}$ are the neuronal activity and
the threshold of the $i$th neuron respectively, $w_{ij}$ is
the element of the connection matrix which stands for the coupling strength 
from the $j$th to the $i$th neuron and $f_{T}(\cdot)$ is a
monotonically increasing function ($0 \leq
f_{T}(\cdot) \leq 1$) with a parameter $T$.
Multiplying $w_{ki}(t+1)$ to both sides of equation (1) and
introducing a new variable
$Y_{ki}(t)=w_{ki}(t)\sigma_{i}(t)$, equation (1)
can be rewritten as
\begin{equation}
Y_{ki}(t+1)=w_{ki}(t+1)f_{T}(\sum_{j=1}^{N}Y_{ij}(t)-h_{i}).
\end{equation}
Here $Y_{ij}$ describes the flow rate of impulse
transmission
from the $j$th to the $i$th neuron.
The variable $Y_{ij}$ represents the strength of the impulse flow 
qualitatively and equation (2) describes the essential effect
of the neuro-flow, but the dynamics of the flow variable $Y_{ij}$
given by equation (2) cannot be uniquely determined.
In order to recover the deterministic equation of the flow
variable $Y_{ij}$, let us replace the term $w_{ki}(t+1)$ by $w_{ki}(t)$ in
equation (2).  Thus, the Markovian equation of the flow variable 
$X_{ij}$ instead of $Y_{ij}$ is obtained
\begin{equation}
X_{ki}(t+1)=w_{ki}(t)f_{T}(\sum_{j=1}^{N}X_{ij}(t)-h_{i}).
\end{equation}
Since our purpose is to elucidate the effect of the neuro-flow, both
equation (2) and equation (3) should be the starting point
for that purpose.  But
from the view point of the consistent description of the system, we
consider that equation (3) is more natural than equation (2).
For the sake of simplicity, the dynamics
described by equation (1) is
called the pattern dynamics, on the other hand, the dynamics given
by equation (3) is known as
the neuro-flow dynamics in this paper.
We rewrite equation (3) using $X_{ki}(t)=w_{ki}(t-1)\sigma_{i}(t)$, the
pattern dynamics of the variable $\sigma_{i}$
corresponding to equation (3) can be given by
\begin{equation}
\sigma_{i}(t+1)=f_{T}(\sum_{j=1}^{N}w_{ij}(t-1)\sigma_{j}(t)-h_{i}).
\end{equation}
This explains the essential difference between equations (1)
and (4), that is to say, equation (1) is simple
Markovian, but equation (4) is not.  Thus, the flow 
dynamics of equation (3) can be expected to reveal quite complex
aspects in comparison with the pattern dynamics
of equation (1).  To summarize, the neuro-flow dynamics,
introduced by equation (3), is equivalent to the pattern
dynamics with time-delayed effects.\\
\begin{center}
{\bf 3. EXTENSION OF THE HEBB RULE}\\
\end{center}
\noindent
The Hebb rule is often written by
\begin{equation}
w_{ij}(t+1)=Aw_{ij}(t)+C(2\zeta_{i}^{\mu}-1)(2\zeta_{j}^{\mu}-1)
\end{equation}
\noindent
where $A$ is the decreasing rate $C$ is the learning
acquisition rate, and the second term of the right hand
side stands for the learning term generated by the
learning patterns \{$\zeta_{i}^{\mu}$\} ($\zeta_{i}^{\mu}=1$
or $0$ with equal probability and $\mu=1,...,p$).  To store
the learning patterns as local
minima, the learning term $2\zeta_{i}^{\mu}-1$ is used
instead of $\zeta_{i}^{\mu}$.
Here we rewrite $\xi_{i}^{\mu}=2\zeta_{i}^{\mu}-1$ and extend
equation (5) by taking account the flow
effect as follows
\begin{equation}
w_{ij}(t+1)=Aw_{ij}(t)+BX_{ij}(t+1)+C\xi_{i}^{\mu}\xi_{j}^{\mu}.
\end{equation}
\noindent
In this paper, we only discuss the case of $B > 0$
according to the original Hebb's idea.  equation (6) is the
simplest extension which introduces the flow effect of
the nerve transmission.\\
\indent
Dynamical systems given by equations (3) and (6) can be
solved by iteration.  Let 
us show the solution for the special case $h_{i}=0$, where the
initial condition is given by $w_{ij}(0)=0$,
$X_{ij}(0)=0$ and $f_{T}(\cdot)$ is a
monotonically increasing function satisfying
$f_{T}(0)=1/2$, $f_{T}(+\infty)=1$ and
$f_{T}(-\infty)=0$.  
The first two steps in the time course can be obtained
\begin{eqnarray}
X_{ki}(1) & = & w_{ki}(0)f_{T}(\sum_{j=1}^{N}X_{ij}(0))
\nonumber \\
& = & 0 \nonumber \\
w_{ki}(1) & = &
Aw_{ki}(0)+BX_{ki}(1)+C\xi_{k}^{1}\xi_{i}^{1} \nonumber \\
& = & C\xi_{k}^{1}\xi_{i}^{1},
\end{eqnarray}
\begin{eqnarray}
X_{ki}(2)&=&w_{ki}(1)f_{T}(\sum_{j=1}^{N}X_{ij}(1))
\nonumber \\
&=&\frac {C}{2}\xi_{k}^{1}\xi_{i}^{1} \nonumber \\
w_{ki}(2)&=&Aw_{ki}(1)+BX_{ki}(2)+C\xi_{k}^{2}\xi_{i}^{2}
\nonumber \\
&=&C\{(A+
\frac{B}{2})\xi_{k}^{1}\xi_{i}^{1}+\xi_{k}^{2}\xi_{i}^{2}\}.
\end{eqnarray}
Here we consider two cases; One is the special case that the value
$\xi_{i}^{\mu}$ of each pattern takes $\pm 1$ randomly with the
constraint $\sum_{j=1}^{N}\xi_{j}^{\mu} \rightarrow
0$ as $N \rightarrow \infty$ (Case 1).  The other is the ordinary
case where the value $\xi_{i}^{\mu}$ takes $\pm 1$ randomly, i.e.
$\sum_{j=1}^{N}\xi_{j}^{\mu} \sim
O(\sqrt{N})$ as $N \rightarrow \infty$ (Case 2).\\

\noindent
{\bf 3.1. Case 1}\\
If we assume the value $\xi_{i}^{\mu}$ of each pattern
takes $\pm 1$ under the constraint that the pattern
fluctuations are negligibly small in the limit of large
system size N (i.e. $\sum_{j=1}^{N}\xi_{j}^{\mu} \rightarrow
0$ as $N \rightarrow \infty$),
the solutions of equations (3) and (6) for the third time
step can be analytically 
obtained 
\begin{eqnarray}
X_{ki}(3)&=&w_{ki}(2)f_{T}(\sum_{j=1}^{N}X_{ij}(2))
\nonumber \\
&=&C\{(A+\frac{B}{2})\xi_{k}^{1}\xi_{i}^{1}+\xi_{k}^{2}\xi_{i}^{2}\}
f_{T}(\frac {C}{2}\xi_{i}^{1}\sum_{j=1}^{N}\xi_{j}^{1})
\nonumber \\
&=&\frac{C}{2}\{(A+\frac{B}{2})\xi_{k}^{1}\xi_{i}^{1}+\xi_{k}^{2}\xi_{i}^{2}\}
\nonumber \\
w_{ki}(3)&=&Aw_{ki}(2)+BX_{ki}(3)+C\xi_{k}^{3}\xi_{i}^{3} 
\nonumber \\
&=&C\{(A+ 
\frac{B}{2})^{2}\xi_{k}^{1}\xi_{i}^{1}+(A+ 
\frac{B}{2})\xi_{k}^{2}\xi_{i}^{2}+\xi_{k}^{3}\xi_{i}^{3}\}.
\end{eqnarray}
After the iteration of $p$ patterns,
the solutions are
\begin{eqnarray}
X_{ki}(p)&=&\frac{C}{2}\sum_{\mu=1}
^{p-1}(A+\frac{B}{2})^{(p-1)-\mu}\xi_{k}^{\mu}\xi_{i}^{\mu}
\nonumber \\
w_{ki}(p)&=&C\sum_{\mu=1}^{p}(A+
\frac{B}{2})^{p-\mu}\xi_{k}^{\mu}\xi_{i}^{\mu}.
\end{eqnarray}
\noindent
For the case of the normal Hebb rule (i.e. $B=0$) the last few
learning
patterns ($\{\xi_{i}^{\mu}\}, \mu \approx p$) can be
well stored
due to the condition $A < 1$.  But when the flow
terms are introduced, i.e.
$(A + B/2) > 1$, the first few learning patterns
($\{\xi_{i}^{\mu}\}, \mu \approx 1$) determine the connection
matrix $w_{ij}$ dominantly.  As a result, the other
learning patterns cannot be stably stored.  
We will explain this point later by numerical simulation.\\

\noindent
{\bf 3.2. Case 2}\\
On the other hand, in the case of the more general condition where we
assume that the value $\xi_{i}^{\mu}$ of each pattern
takes $\pm 1$ randomly, the solutions of equations (3) and (6)
for the third time step change as follows:
\begin{eqnarray}
X_{ki}(3)&=&w_{ki}(2)f_{T}(\sum_{j=1}^{N}X_{ij}(2))
\nonumber \\
&=&C\{(A+\frac{B}{2})\xi_{k}^{1}\xi_{i}^{1}+\xi_{k}^{2}\xi_{i}^{2}\}f_{T}(\frac
{C}{2}\xi_{i}^{1} \times \pm O(\sqrt{N}))
\nonumber \\
&=&C\{(A+\frac{B}{2})\xi_{k}^{1}\xi_{i}^{1}+\xi_{k}^{2}\xi_{i}^{2}\}\frac{1 
\pm \xi_{i}^{1}}{2}
\nonumber \\
w_{ki}(3)&=&Aw_{ki}(2)+BX_{ki}(3)+C\xi_{k}^{3}\xi_{i}^{3} 
\nonumber \\
&=&C\{[(A+\frac{B}{2})\xi_{k}^{1}\xi_{i}^{1}+\xi_{k}^{2}\xi_{i}^{2}]
[A+\frac{B}{2}(1 \pm \xi_{i}^{1})]
+\xi_{k}^{3}\xi_{i}^{3}\}.
\end{eqnarray}
After the iteration of $p$ patterns,
the solutions are
\begin{eqnarray}
X_{ki}(p)&=&C\{(A+\frac{B}{2})\xi_{k}^{1}\xi_{i}^{1}
(\alpha_{i}^{\pm 1})^{p-3}
+\sum_{\mu=2}^{p-1}(\alpha_{i}^{\pm 1})^{(p-1)-\mu}\xi_{k}^{\mu}\xi_{i}^{\mu}\} 
\frac{1 \pm \xi_{i}^{1}}{2}
\nonumber \\
w_{ki}(p)&=&C\{(A+\frac{B}{2})\xi_{k}^{1}\xi_{i}^{1}
(\alpha_{i}^{\pm 1})^{p-2}
+\sum_{\mu=2}^{p}(\alpha_{i}^{\pm 1})^{p-\mu}\xi_{k}^{\mu}\xi_{i}^{\mu}\}
\end{eqnarray}
where $\alpha_{i}^{\pm 1}=A+B(1 \pm \xi_{i}^{1})/2$.  Thus, in the case of 
the random learning patterns without constraints, the connection
matrix is determined by
$\alpha_{i}^{\pm 1}$ instead of ($A+B/2$).  For the case of the
normal Hebb rule ($B=0$), the last few learning patterns can be well 
stored since $A<1$, and the matrix element $w_{ij}(p)$ of
equation (12)
becomes the same as that of equation (10).
However, when $B>0$, $\alpha_{i}^{\pm 1}$ takes $(A+B)$ or $A$.  This
implies that the 
last few learning patterns can be well stored for $(A+B)<1$, but for 
$(A+B)>1$ the
first few learning patterns can be stored.  We will explain this
point later by numerical simulation.\\
\begin{center}
{\bf 4. SLEEP IN NEURO-FLOW DYNAMICS}\\
\end{center}
\noindent
Now let us further investigate equation (6) under the
condition of no learning term, i.e. $C=0$. When the external stimuli stop,
the neural network evolves autonomously according to the neuro-flow
dynamics equations (3) and (6) with $C=0$.  This situation
can be compared with the state where the central nervous 
systems are sleeping, that is, 
the activity of brain is
self-organized only by the internal information.  The
time evolution of $w_{ij}$ 
during sleep can be written as
\begin{eqnarray}
w_{ki}(t+1)&=&Aw_{ki}(t)+BX_{ki}(t+1) \nonumber \\
&=&w_{ki}(t)[A+Bf_{T}(\sum_{j=1}^{N}X_{ij}(t)-h_{i})].
\end{eqnarray}
\noindent
In the case of the ordinary Hebb rule ($B = 0$), $w_{ij}(t)$
vanishes when $t \rightarrow \infty$ because of
$A<1$, but in the case of $B > 0$, $w_{ij}(t)$ 
does not always decrease.  In other words, sleep in
the neuro-flow dynamics brings about two quite different
effects; one is forgetting of the memory, the other is
reinforcement of the memory.  The general criterion
for forgetting or reinforcement is derived from
equation (13):
\begin{equation}
\begin{array}{lll}
\displaystyle{w_{ki}(t+1) \geq w_{ki}(t)}; & \quad\mbox{for} &
\displaystyle{f_{T}(\sum_{j=1}^{N}X_{ij}(t)-h_{i}) \geq
\frac{1-A}{B}} \\
\displaystyle{w_{ki}(t+1) < w_{ki}(t)}; &  \quad\mbox{for} &
\displaystyle{f_{T}(\sum_{j=1}^{N}X_{ij}(t)-h_{i}) <
\frac{1-A}{B}}.
\end{array}
\end{equation}
We discuss the time evolution of $X_{ki}(t)$ and $w_{ki}(t)$ in the
sleep condition for
Case 1 and Case 2 as mentioned before.\\

\noindent
{\bf 4.1. Case 1}\\
In the case of the random patterns with the constraint
$\sum_{j=1}^{N}\xi_{j}^{\mu} \rightarrow
0$ as $N \rightarrow \infty$, 
after the learning obtained by equation (10) is finished,
the addition of the $l$ times sleep induces the following solutions:
\begin{eqnarray}
X_{ki}(p+l)&=&\frac{C}{2}\sum_{\mu=1}^
{p}(A+\frac{B}{2})^{(p-1)+l-\mu}\xi_{k}^{\mu}\xi_{i}^{\mu}
\nonumber \\
w_{ki}(p+l)&=&C\sum_{\mu=1}^{p}(A+
\frac{B}{2})^{p+l-\mu}\xi_{k}^{\mu}\xi_{i}^{\mu}.
\end{eqnarray}
Here in the case of $(A + B/2) > 1$, the
matrix elements $w_{ij}$ corresponding to the
first few learning patterns can be much
amplified after sleep.  This is nothing but the
reinforcement of the primacy effect [8] by sleep.  On
the other hand, every learning pattern is vanishing for
$(A + B/2) < 1$.\\

\noindent
{\bf 4.2. Case 2}\\
In the case of random patterns without constraints, 
the addition of the $l$ times sleep after the learning of
equation (12) induces the following solutions
\begin{eqnarray}
X_{ki}(p+l)&=&C\{(A+\frac{B}{2})\xi_{k}^{1}\xi_{i}^{1}
(\alpha_{i}^{\pm 1})^{p+l-3}
+\sum_{\mu=2}^{p-1}(\alpha_{i}^{\pm 1})^{(p-1)+l-\mu}\xi_{k}^{\mu}\xi_{i}^{\mu}\} 
\frac{1 \pm \xi_{i}^{1}}{2}
\nonumber \\
w_{ki}(p+l)&=&C\{(A+\frac{B}{2})\xi_{k}^{1}\xi_{i}^{1}
(\alpha_{i}^{\pm 1})^{p+l-2}
+\sum_{\mu=2}^{p}(\alpha_{i}^{\pm 1})^{p+l-\mu}\xi_{k}^{\mu}\xi_{i}^{\mu}\}.
\end{eqnarray}
Here every learning pattern will be erased for
$\alpha_{i}^{\pm 1}<1$, but the
first few learning patterns can be amplified for
$\alpha_{i}^{\pm 1}>1$; the condition of the reinforcement of the
primacy effect will be given by $(A+B)>1$.\\
\begin{center}
{\bf 5. NUMERICAL SIMULATION}\\
\end{center}
\noindent
Let us now numerically investigate the neuro-flow
dynamics with the extended Hebb rule.  
The values of 
the parameters used in this paper are fixed as follows:
$N=100$, $p=10$, $h_{i}=0$; the
orthogonal learning patterns $\{\xi_{i}^{\mu}\}$s are
selected from the randomly produced patterns
and they satisfy $\sum_{j=1}^{N}\xi_{j}^{\mu} \approx 0$
and $\sum_{j=1}^{N}\xi_{j}^{\mu}$ is set to $0$
for Case 1.  $f_{T}(x)$ is
fixed as $f_{T}(x)=1/(1+exp(-x/T))$ with $T=0.05$.  The
time development is performed by the synchronous updating of
equations (3) and (6) under the initial conditions of
$w_{ij}(0)=0$ and $X_{ij}(0)=0$.\\

\noindent
{\bf 5.1. Case 1}\\
The connection matrix $w_{ij}(p)$ were determined
after the learning process mentioned above.  The basin
structure corresponding to the fixed matrix $w_{ij}(p)$ was searched by the 
Monte-Carlo simulations, i.e. initial values of the
test 
flow $X_{ij}(0)$s in equation (3) were distributed randomly in the
range $-1 \leq X_{ij}(0) \leq 1$.  The statistical
properties of the basin structure can be represented by
the rank-size relation of each pattern; the
size is defined by the volume of the each basin and the rank of
each basin is defined in accordance with the order of
largeness of the basin volume.  For example,
rank 1 corresponds to the largest basin and rank 
2 to the second largest basin, etc.  Figure 1 shows the
relation between rank $n$ and the basin volume $V$
for the case of  $A=1$ and $B=0$.  Here the filled circles
stand for the learning patterns and the crosses for the
spurious states.  As is shown in Fig. 1, there exist a
number of basins for spurious states and they obey
\begin{equation}
V \propto n^{-D} \hspace {1.0cm} (D=0.5582152)
\end{equation}
where $D$ is the Zipf index.  However all learning
patterns have
a larger basin volume than that of spurious states, and this
clearly shows that the
neuro-flow dynamics proposed here has the learning
ability in the same level as the Hopfield model of
neural networks.  Furthermore, the rank-size relation
obeys an inverse power law.  This is also
similar to the result that was reported in a previous paper
treating the Hopfield model \cite{la9,la10}.\\
\indent
When the flow term is
introduced, the basin structure is drastically changed
as is shown in Figs 2 and 3.  
Figure 2 reveals the flow effect on the learning process 
in the $(A, B)$ parameter
space.   The number of stably stored learning patterns
$M (M \leq p)$ 
is shown in the two-dimensional $(A, B)$ parameter space
in gray-scale.
The axis of $B=0$ 
corresponds to the case of the normal Hebb rule. All
learning patterns are stably stored around
$(A+B/2)=1$ and the
number of stably stored learning patterns $M$ decreases
almost monotonically as the distance from the line
$(A+B/2)=1$ increases.
Figure 3 shows the change of the basin volume of each
learning pattern $(\mu = 1, 2, ..., 10)$ 
for the various values of $B$ at $A = 0.9$.  We can
see that
when the value of $B$ increases, the basin volume for the
first few learning 
patterns $(\mu = 1, 2, 3)$ become dominant for $B\,\,\,\vphantom{(}\lower4pt\hbox{$\stackrel{ \displaystyle {>}}
{\sim}$}\,\,\,  0.2$ and the basin for 
spurious states gets smaller (to less than $10 \%$).  This result clearly
reveals the primacy effect of the memory and is
consistent with the theoretical prediction derived from 
equation (10).  On the other hand, for the case of $B \,\,\,
\vphantom{(}\lower4pt\hbox{$\stackrel{ \displaystyle {<}} {\sim}$}\,\,\,
0.2$, the basin volume for the last few learning
patterns $(\mu = 8, 9, 10)$ become dominant.
In other words, a transition occurs from the 
recency effect \cite{la8} to the primacy effect at the critical
point $B \simeq 0.2$.\\
\indent
Next, further iteration after the learning $(p =
10)$ by the neuro-flow dynamics without 
the learning term was
performed numerically using equations (3) and (6) with 
$l = 1000$.  
Figures 4 and 5 show the results 
of the basin structures and the phase diagram.
Comparing Fig. 4 with Fig. 2, one can see that
memories are completely erased after sleep in the region 
of $(A+B/2) < 1$.  Figure 5
shows the change of the basin volume for each
learning pattern, where the value of parameter $A$ is the 
same as in Fig. 3.  The basin volume for the first few
learning patterns $(\mu = 1, 2, 3)$ increases much more
sharply and the spurious basin becomes
smaller very quickly for $B \geq 0.2$.  On the
other hand, in the case of $B < 0.2$, the basin
volume for all learning patterns is vanishing.  This shows that
the primacy effect discussed in Fig. 3 is much more
emphasized due to the sleep.
Figure 5 appears to suggest that
sleep reinforces and selects memories which were
learned during awakeness, and that a lot of spurious
memories are washed out during sleep.
Actually the system approaches steady states after long
sleep when the effects of the flow increases and the number
of spurious memories becomes less than 10\%.\\

\noindent
{\bf 5.2. Case 2}\\
For the case of random patterns without constraints, the same
numerical simulation was
performed and quite different property was elucidated.\\
\indent
Figure 6 shows the relation between rank $n$ and the basin
volume $V$ for the case of $A=1$ and $B=0$, and spurious states obey 
\begin{equation}
V \propto n^{-D} \hspace {1.0cm} (D=0.5582152).
\end{equation}
Here the learning ability of neuro-flow dynamics is clearly
shown again by the fact that all the learning patterns have
a larger basin
volume than that of spurious states.  As was expected from the 
theoretical analysis, the slope of Fig. 1 and that of Fig. 6 are
completely same.\\
\indent
The flow effect on the learning process is shown in Fig. 7 and it 
is quite different from Fig. 2.  The line $(A+B)=1$ clearly
separates two region: $(A+B)<1$ and $(A+B)>1$.  In $(A+B)<1$, all
the learning 
patterns are stably stored at $A \approx 1$, in which $(A+B) \approx 
A$, and gradually decrease as $A$ decreases.  
However, the number of stably stored patterns
shows a complex structure depending on the combination of the
value of $(A+B)$ and $A$.  On the other hand, in $(A+B)>1$
only one pattern is stored successfully, and this can be understood
from equation (12).  Since $\alpha_{i}^{\pm 1}$ takes $(A+B)$ or $A$
according to the value $\xi_{i}^{1}$ and since $(A+B)>1$ and
$A<1$, all the terms
$\xi_{k}^{\mu}\xi_{i}^{\mu}$ ($\mu=2, 3, \cdots, p$) is totally
disturbed and cannot be stably stored. \\
\indent
Figure 8 shows the change of the basin volume of each learning
pattern for the various values of $B$ at $A=0.9$.  We can clearly see 
the drastic change from the recency effect for
$B\,\,\,\vphantom{(}\lower4pt\hbox{$\stackrel{ \displaystyle {<}}
{\sim}$}\,\,\,  0.1$ to the
primacy effect for $B\,\,\,\vphantom{(}\lower4pt\hbox{$\stackrel
{ \displaystyle {>}}{\sim}$}\,\,\,  0.1$ occurs.  It is also
surprising that all initial flows converge to the first learning
pattern ($\mu=1$) for 
$B>0.1$ and all spurious states are washed out.\\
\indent
Next, the further iteration after learning ($p=10$) was performed 
in the sleep condition with $l=1000$.  Figures 9 and 10 show the
results of the basin structure and the phase diagram.  As was seen in 
Fig. 4, the erasure of memories is also observed for $(A+B)<1$ in
Fig. 9.  Figure 10 shows the change of the basin volume for each
learning pattern, where the value of parameter $A$ is the same as in 
Fig. 8.  We can see that a much sharper change than Fig. 8 occurs at
$B=0.1$, that is, for $B<0.1$ all learning patterns are erased, and
for $B>0.1$ the first learning pattern ($\mu=1$) becomes dominant.
It should be also pointed out that there are no spurious states in
$B>0.1$.\\
\indent
The above results indicate that sleep amplifies the primacy
effect in the case of random patterns.  Therefore, together with the
results of Case 1, it can be suggested that sleep reinforces and 
selects the memories which were learned during awakeness, and that
many spurious memories are erased 
during the sleep.\\
\begin{center}
{\bf 6. SUMMARY}\\
\end{center}
We have proposed a neuro-flow dynamics with the 
extended Hebb rule to deal with the autonomy of the
neural systems and have shown that the neuro-flow
dynamics is equivalent to the pattern dynamics with a
certain time delayed connection.  Two essential new
features are derived from the neuro-flow dynamics.  One is
the reinforcement of the primacy effect of the
memory, the other is the striking effect of sleep
which washes out the spurious memories.  The latter
point is quite similar to the idea that was mentioned
by Click concerning the function of dreaming in
sleep \cite{la11}.  In this connection, the mechanism of the memory
palimpsest, which was studied by Nadal et al. \cite{la12},
in the future must be discussed
in comparison with our neuro-flow dynamics model.\\
\indent
The analysis in this paper is only limited to the case only for
random learning patterns and their special case.  The extension to
the non-random
patterns will be discussed in a forthcoming paper together
with
the generalization of the neuro-flow dynamics model \cite{la13}.\\

\noindent
Acknowledgements -- This work was partially supported by a
Grant-in-Aid for Encouragement of Young Scientists from the Ministry
of Education, Science and Culture.  One of the authors
(M. T.) appreciates the financial support of the Japan
Society for the Promotion of Science for Japanese Junior
Scientists.\\

\newpage

\noindent
Fig. 1. The rank-size relation of Case 1 for the basin volume
obtained by Monte-Carlo simulation with 10000 randomly
chosen initial conditions.  The parameters are
$A=1.0$, $B=0.0$ and 
$C=1.0$.  The filled circles correspond to the
learning patterns and the crosses to spurious states.
The solid line shows the fit of the spurious states by the
least-squares method and the slope is $-0.5582152$.\\

\noindent
Fig. 2. Formation of the basin for the 10 random learning
patterns with the constraint
$\sum_{j=1}^{N}\xi_{j}^{\mu}=0$.  $A$ and $B$ are varied
between 0 and 1 with 0.01
steps.  The gray-scaled $M (M \leq 10)$ stands for the
number of stably stored learning patterns.  In the
region of $(A+B/2) > 1$, the first few learning
patterns $(\mu = 1, 2, 3)$ are stable, but in the region
of $(A+B/2) < 1$, the last few learning patterns
$(\mu = 8, 9, 10)$ are stable.  The result is
successfully explained by equation (10).\\

\noindent
Fig. 3. The basin volume for each learning pattern $(\mu
= 1, 2, ..., 10)$ 
obtained by Monte-Carlo simulation with 1000 randomly
chosen initial conditions.  The parameters are $A=0.9$
and $C=1.0$.  The index $\mu$ is
written at the curve which corresponds to the recency
or the primacy effect.\\

\noindent
Fig. 4. Formation of the basin for the 10 random learning
patterns with the constraint $\sum_{j=1}^{N}\xi_{j}^{\mu}=0$ 
after sleep; $p = 10$ and $l = 1000$.  $A$ and $B$ are varied
between 0 and 1 with 0.01 steps.  All learning
patterns are only stably stored around the line
$(A+B/2) = 1$ which is consistent with
equation (13).  In the region of $(A+B/2) < 1$, 
memories for all learning patterns were completely washed
out by sleep and in the region of $(A+B/2) > 1$, the first few
learning patterns $(\mu = 1, 2, 3)$ remain very strongly 
even after a long sleep.\\

\noindent
Fig. 5. The basin volume for each learning pattern
after sleep; $p = 10$ and $l=1000$.  The basin volume for
the first few learning patterns increases very sharply when the
value of $B$ increases.  This implies reinforcement
of the primacy effect by sleep.  The parameters are $A=0.9$ and
$C=1.0$.  The index $\mu$ is
written at the curve which corresponds to the primacy effect.\\

\noindent
Fig. 6. The rank-size relation of Case 2 for the basin volume
obtained by Monte-Carlo simulation with 10000 randomly
chosen initial conditions.  The parameters are
$A=1.0$, $B=0.0$ and 
$C=1.0$.  The filled circles correspond to the
learning patterns and the crosses to spurious states.
The solid line is the fit of the spurious states by the
least-squares method and the slope is $-0.5582152$.  This
figure is the same as Figure 1 because the matrix element
$w_{ij}(p)$ of equations (10) and (12)
becomes equal for $B=0$.\\

\noindent
Fig. 7. Formation of the basin for the 10 random learning
patterns.  $A$ and $B$ are varied between 0 and 1 with 0.01
steps.  In the
region of $(A+B) > 1$, only the first learning
pattern $(\mu = 1)$ is stable, but in the region
of $(A+B) < 1$, all learning patterns are stably stored
at $A \approx 1$ and decrease as $A$ decreases.  The result is
explained qualitatively well by equation (12).\\

\noindent
Fig. 8. The basin volume for each learning pattern $(\mu
= 1, 2, ..., 10)$  
obtained by Monte-Carlo simulation with 1000 randomly
chosen initial conditions.  The parameters are $A=0.9$
and $C=1.0$.  The index $\mu$ is
written at the curve which corresponds to the recency
or the primacy effect.\\

\noindent
Fig. 9. Formation of the basin for the 10 random learning
patterns after sleep; $p = 10$ and $l = 1000$.  $A$ and $B$ are varied
between 0 and 1 with 0.01 steps.  In the region
of $(A+B) < 1$, 
memories for all learning patterns were completely washed
out by sleep and in the region of $(A+B) > 1$, the first
learning pattern $(\mu = 1)$ remains very strong
even after a long sleep.\\

\noindent
Fig. 10. The basin volume for the 10 learning patterns
after sleep; $p = 10$ and $l=1000$.  The basin volume for
the first learning pattern increases very sharply when the
value of $B$ increases.  This implies reinforcement
of the primacy effect by sleep.  The parameters are $A=0.9$ and
$C=1.0$.  The index $\mu$ is
written at the curve which corresponds to the primacy effect.
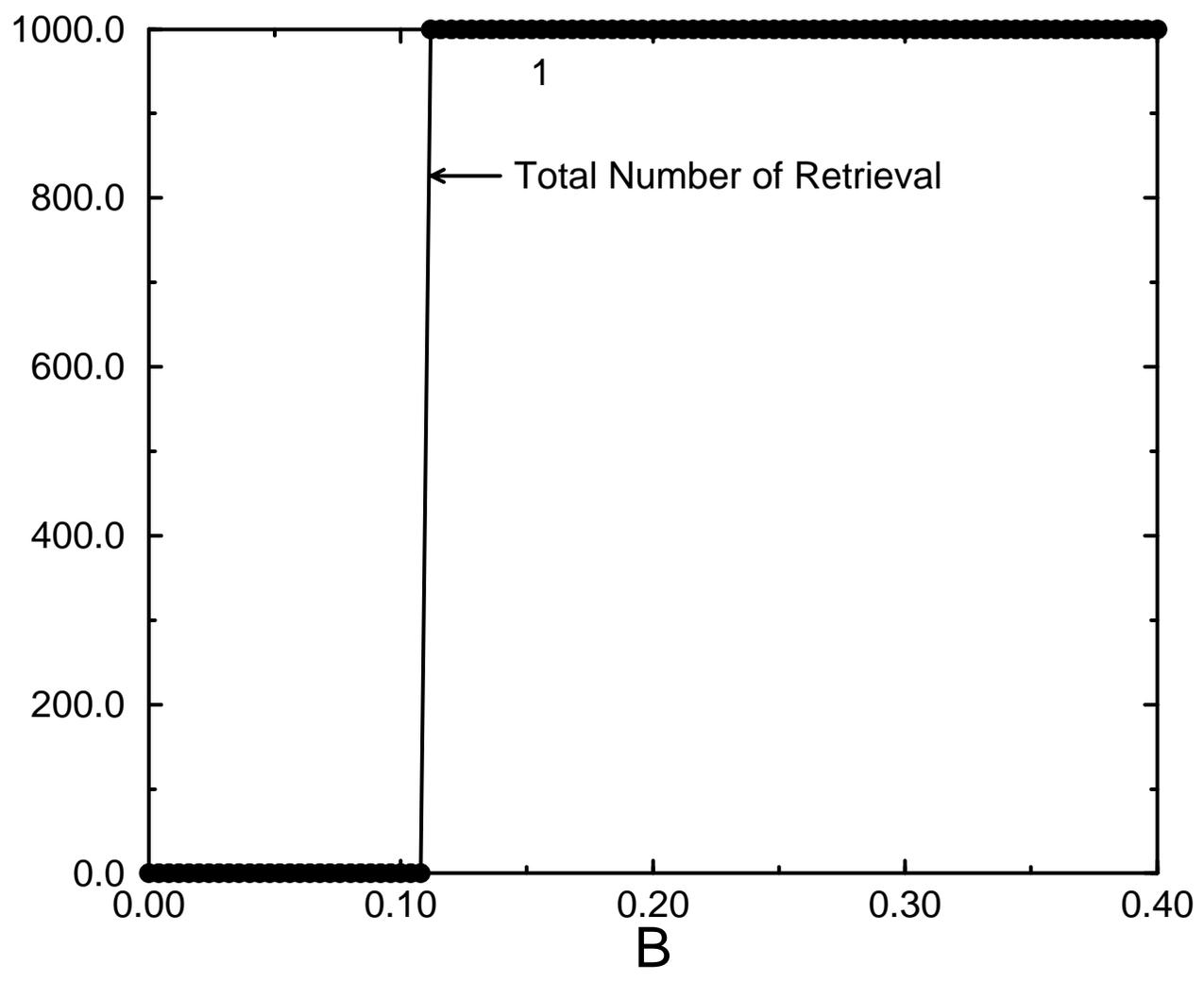
\end{document}